\documentclass{IEEEtran}
\usepackage{cite}
\usepackage{amsmath,amssymb,amsfonts}
\usepackage{algorithmic}
\usepackage{graphicx}
\usepackage{textcomp}
\usepackage{algorithm,algorithmic}
\def\BibTeX{{\rm B\kern-.05em{\sc i\kern-.025em b}\kern-.08em
    T\kern-.1667em\lower.7ex\hbox{E}\kern-.125emX}}
\begin{document}
\title{Dual Detection Framework for Faults and Integrity Attacks in Cyber-Physical Control Systems}
\author{Xixing Xue, Dong Shen \IEEEmembership{Senior Member, IEEE}, Steven X. Ding, and Dong Zhao \IEEEmembership{Senior Member, IEEE}
	\thanks{Xixing Xue and Dong Zhao are with the School of Cyber Science and Technology, Beihang University, Beijing 100191, China (e-mail: by2339213@buaa.edu.cn; dzhao@buaa.edu.cn).}
	\thanks{Dong Shen is with the School of Mathematics, Renmin University of China, Beijing 100872, China (e-mail: dshen@ieee.org).}
	\thanks{S. X. Ding is with the Institute for Automatic Control and Complex Systems (AKS), University of Duisburg-Essen, 47057 Duisburg, Germany (e-mail: steven.ding@uni-due.de).}
}
\maketitle

\begin{abstract}
	Anomaly detection plays a vital role in the security and safety of cyber-physical control systems, and accurately distinguishing between different anomaly types is crucial for system recovery and mitigation.  This study proposes a dual detection framework for anomaly detection and discrimination. By leveraging the dynamic characteristics of control loops and the stealthiness features of integrity attacks, the closed-loop stealthiness condition is first derived, and two dedicated detectors are designed and deployed on the controller side and the plant side, respectively, enabling joint plant fault and cyber attack detection. Moreover, by jointly analyzing the residual response of the two detectors corresponding to different anomalies, it is proved that the proposed method can distinguish between faults and integrity attacks due to the detectors' individual residual spaces. According to the detector's residual space, the fault and attack detection performance is further improved by a two-stage optimization scheme. Simulation results validate the effectiveness of the proposed approach.
\end{abstract}


\section{Introduction}
Cyber-physical systems (CPSs) integrate computational components with physical processes through communication networks, enabling precise sensing, control, and coordination. Such systems significantly improve  operation efficiency and intelligence. However, as CPSs grow in scale and complexity, their security and safety become increasingly critical. One major challenge arises from system faults, such as hardware degradation and sensor failures\cite{article1}. In addition, network communication also exposes these systems to cyber threats, especially integrity attacks\cite{article2}. 

Timely and accurate anomaly detection is crucial for fault isolation and system recovery, thereby ensuring the safe and stable operation of CPSs\cite{article3}.  Over the past few decades, fault detection has been extensively studied, and a variety of model-based methods have been developed\cite{article4}, including observer-based approaches\cite{article5,article6,article7}, parity space methods\cite{article8,article9}, and active fault diagnosis\cite{article10,article11,article12}. In recent years, driven by the rapid progress of artificial intelligence, numerous learning-based methods have also been proposed\cite{article13,article14,article15}. These techniques have demonstrated high effectiveness in detecting system faults and have been widely applied in various engineering systems.

Despite the maturity of fault detection techniques, they exhibit inherent limitations in addressing integrity attacks. Whereas faults typically result from natural failures or component degradation, integrity attacks are intentionally designed to manipulate data and bypass fault detectors, posing significant challenges to anomaly detection in CPSs. Although various attack detection methods, such as model-based approaches\cite{article16,article18,article20} and learning-based approaches \cite{article22,article23,article24}, have been developed, detecting integrity attacks remains an open problem to be addressed. By exploiting system knowledge, attackers can design injected data that satisfies stealthiness conditions\cite{article26,article28}, enabling them to evade not only traditional fault detection mechanisms but also many existing protection schemes. Moreover, most existing studies treat attack detection and fault diagnosis separately, while in practical applications, both are often required simultaneously.

Fundamentally, the essence of anomaly detection lies in not only detecting abnormal behavior but also differentiating between various types of anomalies. In CPSs, both faults and attacks can disrupt normal system operation, yet they differ significantly in their mitigation strategies. Failure to effectively discriminate between faults and attacks may lead to inappropriate corrective actions, potentially exacerbating system damage. Despite its critical importance, anomaly distinction has often been overlooked in existing research. Some studies focus solely on differentiating various fault types without considering attack scenarios\cite{article33,article34,article35}, while some research that addresses both faults and attacks frequently incurs system modifications or complex data processing\cite{article36,article37,article38,article39}.

Building on this background, our study returns to the core objective of anomaly detection: identifying and distinguishing between faults and attacks. The study in \cite{article28} summarizes stealthy attacks such as zero-dynamics attacks and covert attacks as kernel attacks. When attacks satisfy the kernel space condition of system plants, they remain undetected by fault detectors. Observing from the viewpoint of the entire control-loop, we note that this kernel space property represents a critical distinction between integrity attacks and faults, which can be exploited to design a low-cost and efficient detection and classification mechanism. Detecting such stealthy attacks using only controller-side data is challenging. Effective detection of kernel attacks requires leveraging the closed-loop characteristics of the control system. 
In the view of the action space of anomalies signals, implementing attack detection at the plant side is more effective. This is due to the fact that it is difficult for the attack to simultaneously satisfy the kernel space conditions at both the plant and the controller sides.
This motivates a natural design choice of deploying coordinated detectors on both the controller side and the plant side. The controller can be inherently regarded as a dynamic system, making it possible to verify its input-output data. This enables the detection of kernel attacks that remain undetectable on the controller side by incorporating complementary observations from the plant side. 
Moreover, this detection design does not incur the need for modification of the control system model, the addition of an authorization signal, or the implementation of encryption mechanisms, thus facilitating a more straightforward implementation.

Motivated by the above analysis, we propose a dual detection framework to improve the system’s capability of detecting anomalies, particularly stealthy attacks, by deploying anomaly detectors on both the controller and plant sides. This framework enables the detection of stealthy attacks and the differentiation between faults and attacks by analyzing the differing responses of the two detectors. We further analyze the stealthiness conditions under this framework and formally define them as the closed-loop stealthiness condition.  In addition, we provide an optimal implementation of the framework through a two-stage optimization in which the detectors on both sides are optimally configured and the controller parameters are jointly optimized to improve detection performance.
The main contributions of this study are listed as follows. 
\begin{enumerate}[]
	\item We propose a dual detection framework for system faults and integrity attacks that leverages the characteristics of the closed-loop control system, without incurring control performance loss or requiring plant dynamic modifications. Moreover, the proposed detection framework enables the discrimination between attacks and system faults.
	
	\item  Within the closed-loop control system, we clarify the channels and response spaces of system faults and attacks, respectively, and formulate the closed-loop stealthiness condition for integrity attacks.
	
	\item  We derive a two-stage optimization scheme for the proposed detection framework,  along with a corresponding threshold design. 
\end{enumerate}

This paper is organized as follows. The considered cyber-physical systems and the kernel attack are described in Section \ref{s2}.  Section \ref{s3} presents the core contributions of this study, introducing the dual detection framework and the corresponding closed-loop stealthiness condition. Section \ref{s4} provides the optimal detection design. The effectiveness of these methods is illustrated in Section \ref{s5}. Section \ref{s6} provides concluding remarks.

$\mathbf{Notation}$. For vector $a$, $\Sigma_a$ denotes the covariance matrix of $a$. For a matrix $X$, $X^T$ and $X^{-1}$ are the transpose and inverse of $X$, respectively. $N(\mu,\Sigma)$ denotes the Gaussian distribution with mean $\mu$ and covariance matrix $\Sigma$. $\rho(X)$ is the spectral radius of matrix $X$. $\mathcal{R}^n$ represents $n$-dimensional Euclidean space, and $\mathcal{C}^n$ represents $n$-dimensional complex vector space. $\mathcal{RH}_\infty$ denotes the set of stable and proper transfer functions with bounded $\mathcal{H}_\infty$ norm. $\mathcal{H}_2$ denotes the set of stable and proper transfer functions with finite $\mathcal{H}_2$ norm. For notational simplicity, the time index $k$ and the frequency-domain variable $z$ are omitted in some expressions without loss of clarity. 

\section{Preliminaries}\label{s2}
In this section, we present the CPS model, which includes the plant, controller, anomaly detector, potential cyber attacks, and system faults. We further analyze the limitations of conventional fault detectors in identifying stealthy attacks, particularly the kernel attack.
\subsection{Cyber-Physical Control Systems}
Consider a linear time-invariant (LTI) system: 
\begin{figure}[!t]
	\centerline{\includegraphics[width=\columnwidth]{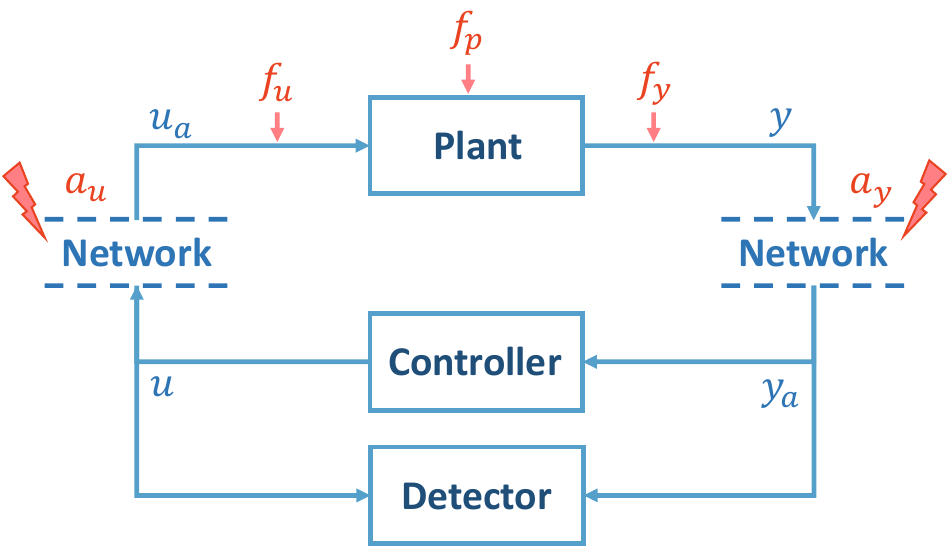}}
	\caption{Cyber-physical control system diagram.}
	\label{model1}
\end{figure}
\begin{equation}
	\begin{cases}x(k+1)=Ax(k)+Bu(k)+\omega(k) \\ y(k)=Cx(k)+Du(k)+\eta(k)
	\end{cases}\label{eq1}
\end{equation}
where $ x(k)\in \mathcal{R}^n$, $y(k)\in \mathcal{R}^p$, and $u(k)\in \mathcal{R}^m$ are system state, measurement, and control signal, respectively. The system noise $\omega(k)\in \mathcal{R}^n$ and measurement noise $\eta(k)\in \mathcal{R}^p$ are independent zero-mean Gaussian noises with covariance matrices $\Sigma_\omega$ and $\Sigma_\eta$, respectively. $x(0)$ is the initial state independent of $\omega(k)$ and $\eta(k)$. $A$, $B$, $C$ and $D$ are system matrices with appropriate dimensions. Without loss of generality, we assume that system \eqref{eq1} is controllable and observable.

We consider observer-based feedback control and the generation of detection residuals for anomaly monitoring. The system observer is
\begin{equation}
	\begin{cases}
		\hat{x}(k+1)=A\hat{x}(k)+Bu(k)+L(y(k)-\hat{y}(k))\\
		\hat{y}(k)=C\hat{x}(k)+Du(k)
	\end{cases}
	\label{eq2}
\end{equation}
where $\hat{x}(k)$ and $\hat{y}(k)$ are the estimates of system state and measurement, respectively. $L$ is the observer gain. The system residual is defined as
\begin{equation}
	r(k)=y(k)-\hat{y}(k)\label{eq3}
\end{equation}   
The feedback control strategy is 
\begin{equation}
	u(k)=F\hat{x}(k)+Qr(k)+\bar{v}(k)\label{eq4}
\end{equation}
where $F$ and $Q$ are controller parameters. $\bar{v}$ denotes the feedforward control component associated with the reference input $v$.

Without attacks, the residual consists of only the estimation error and measurement noise. Considering the Gaussian properties of $\omega$ and $\eta$, attack detection can be achieved by using a Kalman filter with different detection statistics, such as $\chi^2$ test or generalized likelihood ratio (GLR), with the GLR corresponding to the optimal detection statistic.

Given the known matrices $\Sigma_\omega$ and $\Sigma_\eta$, the Kalman gain is calculated by
\begin{equation}
	L_k:=L=APC^T\Sigma_r^{-1}\label{eq5}
\end{equation}
\begin{equation}
	P=APA^T+\Sigma_\omega-L_k\Sigma_rL_k^T\label{eq6}
\end{equation}
\begin{equation}
	\Sigma_r=CPC^T+\Sigma_\eta
	\label{eq7}
\end{equation}

Based on the factor that $r(k)\sim N(0,\Sigma_r)$, the residual-based $\chi^2$ detector with threshold $J_{th}$ is
\begin{equation}
	J(k)=r(k)^T\Sigma_r^{-1}r(k)\overset{H_1}{\underset{H_0}{\gtrless}}J_{th}\label{eq8}
\end{equation}
The $\chi^2$ detector is sensitive to system faults $f_u$, $f_y$, and $f_p$, as these faults typically disturb the Gaussian characteristics of the residual $r(k)$. As shown in Fig. \ref{model1}, $f_u$, $f_y$, and $f_p$ represent actuator faults, sensor faults, and plant faults, respectively. 

$\mathbf{Remark~1.}$
Consider the process model
\begin{equation}
	\begin{cases}
		x(k+1)=Ax(k)+Bu(k)+E_ff(k)+E_dd(k)\\
		y(k)=Cx(k)+Du(k)+F_ff(k)+F_dd(k)
		\label{eq58}
	\end{cases}
\end{equation}
where $d(k)\in \mathcal{R}^{d}$ and $f(k)\in\mathcal{R}^f$ represent the unknown input and fault, respectively. It is proved that the residual generator
\begin{equation}
	\begin{cases}
		\hat{x}(k+1)=A\hat{x}(k)+Bu(k)+L_{opt}(y(k)-\hat{y}(k))\\
		\hat{y}(k)=C\hat{x}(k)+Du(k)\\
		r(k)=V_{opt} (y(k)-\hat{y}(k))\\
		L_{opt}=(AXC^T+E_dF_d^T)V_{opt}^2\\
		V_{opt}=(CXC^T+F_dF_d^T)^{-\frac{1}{2}}
	\end{cases}\label{eq56}
\end{equation}
where $X>0$ solves the Riccati equation
\begin{equation}
	AXA^T-X+E_dE_d^T-L_{opt}(CXC^T+F_dF_d^T)L_{opt}^T=0\notag
\end{equation}
is optimal in the sense of 
\begin{equation}
	\forall \theta\in[0,2\pi], \{L_{opt},V_{opt}\}=\arg \sup_{L~V}\frac{\sigma_i(V\hat{N}_f(e^{j\theta}))}{\|V\hat{N}_d\|_\infty}
\end{equation}
\begin{equation}
	\hat{N}_f(z)=F_f+C(zI-A+LC)^{-1}(E_f-LF_f)	\notag
\end{equation}
\begin{equation}
	\hat{N}_d(z)=F_d+C(zI-A+LC)^{-1}(E_d-LF_d)\notag
\end{equation}
with $\sigma_i(V\hat{N}_f(e^{j\theta}))$ denoting the singular values of $V\hat{N}_f(e^{j\theta})$. $V \hat{N}_f(z)$ and $V\hat{N}_d(z)$ represent the transfer functions from $f$ and $d$ to $r$ , respectively. The solution \eqref{eq56} is called unified solution\cite{article40}. 

With the stochastic system \eqref{eq1}, this Kalman filter-based mechanism is consistent with the unified solution \eqref{eq56} and achieves optimal performance compared with other detection systems\cite{article29}. For deterministic systems with deterministic disturbances, the unified solution \eqref{eq56} can be applied to achieve the same optimality.

\subsection{Kernel Attack Against Fault Detector}
Given the attacked system model
\begin{equation}
	\begin{cases}
		x_a(k+1)=Ax_a(k)+Bu_a(k)+\omega(k)\\
		y_a(k)=Cx_a(k)+Du_a(k)+a_y(k)+\eta(k)
		\label{eq9}
	\end{cases}
\end{equation}
where $u_a(k)=u(k)+a_u(k)$ denotes the attacked control input. $a_u(k)$ and $a_y(k)$ represent the attacks injected into the control signal and the measurement, respectively. The detection system \eqref{eq2}-\eqref{eq3} has been proven incapable of defending a series of stealthy attacks, such as replay attacks, covert attacks, and zero dynamics attacks\cite{article30}. These attacks can be concluded as kernel attack\cite{article28}. To introduce the kernel attack, we firstly discuss the transfer function representation of system \eqref{eq1}-\eqref{eq4}. Given the nominal plant model
\begin{equation}
	y(z)=G_u(z)u(z)\label{eq10}
\end{equation}
where $u(z)\in \mathcal{C}^m$ and $y(z)\in \mathcal{C}^p$ are the plant input and output vectors, respectively. $G_u (z)\in \mathcal{RH}_\infty$ is the transfer function. The left and right coprime factorizations (LCF and RCF) of $G_u(z)$ are given by
\begin{equation}
	G_u(z)=\hat{M}^{-1}(z)\hat{N}(z)=N(z)M^{-1}(z)\label{eq11}
\end{equation}
The state space realizations of the left and right coprime pairs (LCP and RCP) $(\hat{M}(z),\hat{N}(z))$ and $(M(z),N(z))$ are
\begin{equation}
	\begin{cases}
		\hat{M}(z)&=(A-LC, -L, C, I)\\
		\hat{N}(z)&=(A-LC, B-LD, C, D)\\
		M(z)&=(A+BF, B, F, I)\\
		N(z)&=(A+BF, B, C+DF, D)
	\end{cases}\label{eq12}
\end{equation}
where $F$ and $L$ are selected such that $A+BF$ and $A-LC$ are Schur matrices. Correspondingly, there exist $(\hat{X}(z),\hat{Y}(z))$ and $(X(z),Y(z))$ over $\mathcal{RH}_\infty$ so that the so-called Bezout identity holds
\begin{equation}
	\begin{bmatrix}
		X(z)&Y(z)\\-\hat{N}(z)& \hat{M}(z)
	\end{bmatrix}
	\begin{bmatrix}
		M(z)&-\hat{Y}(z)\\N(z)&\hat{X}(z)
	\end{bmatrix}=
	\begin{bmatrix}
		I&0\\0&I
	\end{bmatrix}\label{eq13}
\end{equation}
\begin{equation}
	\begin{cases}
		\hat{X}(z)&=(A+BF, L,C+DF, I)\\
		\hat{Y}(z)&=(A+BF, -L, F, 0)\\
		X(z)&=(A-LC, -(B-LD), F, I)\\
		Y(z)&=(A-LC, -L, F, 0)
	\end{cases}\label{eq14}
\end{equation}
Then, the residual generator can be written as 
\begin{equation}
	r_0(z)=\hat{M}(z)y(z)-\hat{N}(z)u(z)\label{eq15}
\end{equation}
All LTI residual generators can be parameterized by
\begin{equation}
	r(z)=R(z)r_0(z)=R(z)(y(z)-\hat{y}(z))
	\label{eq16}
\end{equation}
where $R(z)\in \mathcal{RH}_\infty$ is the parameterization function.

The feedback controller \eqref{eq2}-\eqref{eq4} can be equivalently written as
\begin{equation}
	u(z)=K(z)y(z)+v(z)
	\label{eq17}
\end{equation}
All stabilizing controllers can be parameterized by Youla parameterization
\begin{align}
	K(z)&=-(X(z)-Q(z)\hat{N}(z))^{-1}(Y(z)+Q(z)\hat{M}(z)) \label{eq18}\\
	&=-(\hat{Y}(z)+M(z)Q(z))(\hat{X}(z)-N(z)Q(z))^{-1}\notag
\end{align}
where the parameter system $Q\in \mathcal{RH}_\infty$.

With the attacked system model \eqref{eq9}, the residual is
\begin{equation}
	r_a(z)=R(z)(\hat{M}(z)y_a(z)-\hat{N}(z)u(z))\label{eq19}
\end{equation}
Then, the kernel attack can be described by Definition 1.

$\mathbf{Definition\ 1}$\cite{article28}. Given system model \eqref{eq10} and the attack detector \eqref{eq15}, an integrity attack is called kernel attack when condition \eqref{eq20} holds.
\begin{equation}
	\begin{bmatrix}
		u(z)\\y_a(z)
	\end{bmatrix}\in \mathcal{K}_P
	\label{eq20}
\end{equation}
where the subspace $\mathcal{K}_P$ is the kernel space of the plant.
\begin{equation}
	\mathcal{K}_P=\{\begin{bmatrix}
		u\\y
	\end{bmatrix}: \begin{bmatrix}
		-\hat{N}&\hat{M}
	\end{bmatrix}\begin{bmatrix}
		u\\y
	\end{bmatrix}=0, \begin{bmatrix}
		u\\y
	\end{bmatrix}\in \mathcal{H}_2\}
	\label{eq21}\notag
\end{equation}

It is easy to prove that kernel attack is stealthy to detector \eqref{eq15}.
\begin{equation}
	\begin{aligned}
		r(z)&=R(z)(y_a(z)-\hat{y}(z))\\&=R(z)\begin{bmatrix}
			-\hat{N}(z)&\hat{M}(z)
		\end{bmatrix}\begin{bmatrix}
			u(z)\\y_a(z)
		\end{bmatrix}
	\end{aligned}
	\label{eq22}\notag
\end{equation}
\begin{equation}
	\begin{bmatrix}
		-\hat{N}(z)&\hat{M}(z)
	\end{bmatrix}\begin{bmatrix}
		u(z)\\y_a(z)
	\end{bmatrix}=0\iff\begin{bmatrix}
		u(z)\\y_a(z)
	\end{bmatrix}\in \mathcal{K}_P
	\label{eq23}\notag
\end{equation}

It is evident that zero-dynamics attacks, covert attacks, and replay attacks fall into the category of kernel attacks\cite{article28}. For simplicity, we consider system \eqref{eq9} with $\omega=0$ and $\eta=0$. A zero-dynamics attack injects malicious inputs $a_u$ without affecting the system output during the detection interval $[k_0,k_0+N]$, thereby remaining undetectable. Attacked output $y_a$ satisfies the condition
\begin{equation}
	y_a(k)=y(k),~\forall k\in [k_0,k_0+N]\notag
\end{equation}
which means $a_u$ satisfies
\begin{equation}
	\hat{N}(z)a_u(z)=0\label{eq55}
\end{equation}
According to \eqref{eq55}, zero-dynamics attacks satisfy the kernel attack condition \eqref{eq20}.
\begin{equation}
	r_0=\begin{bmatrix}
		-\hat{N}& \hat{M}
	\end{bmatrix}\begin{bmatrix}
		u\\y_a
	\end{bmatrix}=\hat{N}a_u=0\notag
\end{equation}

Considering the attacked system model \eqref{eq9}, covert attacks inject $a_u$ into the control signals and $a_y$ into the measurements, respectively, and satisfy
\begin{equation}
	a_y+G_ua_u=0\implies y_a(k)=y(k),~\forall k\in[k_0, k_0+N]\notag
\end{equation}
It follows that covert attacks satisfy
\begin{equation}
	\hat{M}(z)a_y(z)+\hat{N}(z)a_u(z)=0\notag
\end{equation}
Hence, condition \eqref{eq20} holds for covert attacks.
\begin{equation}
	r_0=\begin{bmatrix}
		-\hat{N}& \hat{M}
	\end{bmatrix}\begin{bmatrix}
		u\\y_a
	\end{bmatrix}=\hat{N}a_u+\hat{M}a_y=0\label{eq57}
\end{equation}

Replay attacks are carried out under the assumption that the dynamic system is stable, which leads to
\begin{equation}
	y(k)\approx y(i),~i=1,2,\ldots\notag
\end{equation}
Specifically, the attacker records and stores a sequence of past sensor measurements $y(k_0+i),~i=0,1,\ldots,M$, and subsequently replay them during the time interval $[k,k+N]$, with $k>k_0+M$. Meanwhile, attackers inject $a_u$ into the control signals. The replay attack can be modeled as
\begin{equation}
	\begin{aligned}
		x_a(j+1)&=Ax_a(j)+Bu_a(j)\notag\\
		y_a(j)&=Cx_a(j)+Du_a(j)+a_y(j)\notag\\
		a_y(j)&=y(k_0+j-k)-(Cx(j)+Du_a(j))\notag
	\end{aligned}
\end{equation}
where $j\in[k,k+M]$. As a result, the residual satisfies
\begin{equation}
	r_0(j)=y(k_0+j-k)-(C\hat{x}(j)+Du(j))\approx 0 \notag
\end{equation}
which implies
\begin{equation}
	\begin{bmatrix}
		u \\ y_a
	\end{bmatrix} \in \mathcal{K}_P \notag
\end{equation}
It should be noted that replay attacks signal $a_y(j)$ depend on the system dynamic, causing multiplicative rather than purely additive effects.

Therefore, zero-dynamics attacks, covert attacks, and replay attacks are all kernel attacks, as they keep the system trajectory within the kernel space, making them hard to detect with standard residual-based methods.

\subsection{Problem Formulation}
Kernel attacks exploit the plant dynamics to remain stealthy to fault detectors, posing a critical challenge to closed-loop control systems. This results in difficulty achieving effective detection based solely on controller-side data. Existing methods are also inadequate in distinguishing between faults and attacks, which adversely affects subsequent system recovery and decision-making. The main goal of this study is to develop an effective detection framework to address the above two issues. We note that detecting kernel attacks requires leveraging closed-loop data, since kernel attacks  take effect within the plant’s kernel space.  Moreover, the stealthiness of kernel attacks distinguishes them from faults. Designing a detection mechanism exploiting this feature is the key to achieving anomaly discrimination without increasing system burden. Building on these considerations, we design a stealthy attack detector and discriminator without affecting the control system, altering the plant dynamics, or requiring additional terminal-side encryption and quasi-key synchronization.

\section{Dual Detection Framework}\label{s3}
In this section, we introduce the dual detection framework and corresponding close-loop stealthiness condition under this framework.

\subsection{Dual Detection Framework}\label{s31}
From the perspective of controller-side fault detection, the detector verifies the consistency between system input $u$ and output $y_a$ by comparing the predicted output $\hat{y}$ with the received measurement $y_a$. The discrepancy between the two reflects whether the system is operating under normal condition or system faults. From \eqref{eq15} and \eqref{eq16}, the detector embodies the concept of kernel-space-based fault detection. Its core principle is to examine whether the control input $u$ and the feedback output $y_a$ match the kernel space features.

\begin{figure}[!t]
	\centerline{\includegraphics[width=\columnwidth]{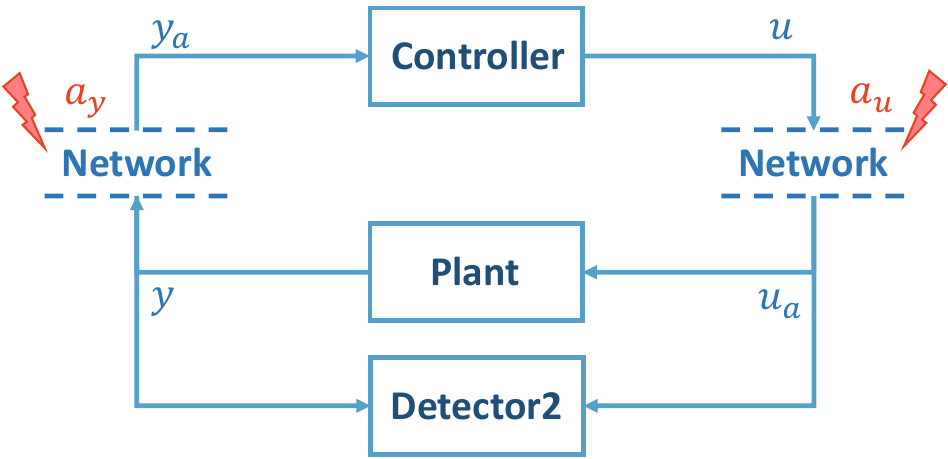}}
	\caption{Plant-side attack detection.}
	\label{model3}
\end{figure}
From the closed-loop perspective, the controller and the plant continuously interact and exchange data. This implies that a similar detection scheme can also be implemented on the plant side. Specifically, by comparing the predicted input $\hat{u}$ with the actual feedback input $u_a$, the consistency between $u_a$ and $y$ can be verified. As shown in Fig. \ref{model3}, this leads to a dual problem of fault detection. For simplicity, we consider the system \eqref{eq1} with $\omega=\eta=0$ in the following analysis. We model the controller as a dynamic system with input $y$ and output $u$.
Considering the linear controller as a dynamic system with transfer function $G_y (z)$.
\begin{equation}
	u_0(z)=G_y(z)y(z)
	\label{eq24}
\end{equation}
\begin{equation}
	u(z)=G_y(z)y(z)+v(z)
	\label{eq25}
\end{equation}
The corresponding state-space function is
\begin{equation}
	\begin{cases}
		\hat{x}(k+1)=\bar{A}\hat{x}+\bar{B}_1y(k)+\bar{B_2}v(k)\\
		u(k)=\bar{C}\hat{x}(k)+\bar{D}_1y(k)+\bar{D}_2v(k)
	\end{cases}
	\label{eq26}
\end{equation}
where $\bar{A}=A+B(F-QC)-LC$, $\bar{B}=[\bar{B}_1,\bar{B}_2]=[L+BQ, B]$,  $\bar{C}=F-QC$, and $\bar{D}=[\bar{D}_1, \bar{D}_2]=[Q,I]$.
The detection scheme can be achieved similarly to \eqref{eq9}-\eqref{eq15}. The LCF and RCF of $G_y(z)$ is denoted as 
\begin{equation}
	G_y(z)=\hat{M}_y^{-1}(z)\hat{N}_y(z)=N_y(z)M_y^{-1}(z)
	\label{eq27}
\end{equation}
\begin{equation}
	\begin{cases}
		\hat{M}_y(z)&=(\bar{A}-L_u\bar{C}, -L_u, \bar{C}, I)\\
		\hat{N}(z)&=(\bar{A}-L_u\bar{C}, \bar{B}-L_u\bar{D}, \bar{C}, \bar{D})\\
		M_y(z)&=(\bar{A}+\bar{B}F_u, \bar{B}, F_u, I)\\
		N_y(z)&=(\bar{A}+\bar{B}F_u, \bar{B}, \bar{C}+\bar{D}F_u, \bar{D})
	\end{cases}
	\label{eq28}
\end{equation}
where $F_u$ and $L_u$ are selected such that $\bar{A}+\bar{B}F_u$ and $\bar{A}-L_u\bar{C}$ are Schur matrices. Then, the controller residual is given as
\begin{equation}
	r_{0,u}(z)=\hat{M}_y(z)u(z)-\hat{N}_y(z)y(z)-\hat{M}_y(z)v(z)
	\label{eq29}
\end{equation}
Similarly to \eqref{eq16}, all LTI residual generators can be parameterized by 
\begin{equation}
	r_u(z)=R_u(z)r_{0,u}(z)\notag
\end{equation}
where $R_u(z)\in \mathcal{RH}_\infty$ is the parameterization function.
Given the system \eqref{eq25} without attacks, we have
\begin{equation}
	r_{0,u}(z)=\hat{M}_y(z)u(z)-\hat{N}_y(z)y(z)-\hat{M}_y(z)v(z)=0
	\label{eq31}
\end{equation}
Consider the attacked controller system
\begin{equation}
	\begin{cases}
		\hat{x}_a(k+1)=\bar{A}\hat{x}_a+\bar{B}_1y_a(k)+\bar{B_2}v(k)\\
		u_a(k)=\bar{C}\hat{x}_a(k)+\bar{D}_1y_a(k)+a_u(k)+\bar{D}_2v(k)
	\end{cases}
	\label{eq32}
\end{equation}
where $y_a(k)=y(k)+a_y(k)$. The residual is generated by
\begin{equation}
	r_{u,a}(z)=R_u(z)(\hat{M}_yu_a(z)-\hat{N}_yy(z)-\hat{M}_y(z)v(z))\label{eq33}
\end{equation}
It is straightforward to give the corresponding stealthiness condition using the same formation as in Definition 1.

For control system \eqref{eq26}, an integrity attack is stealthy to detector \eqref{eq29} if and only if condition \eqref{eq34} holds.
\begin{equation}
	\begin{bmatrix}
		y(z)\\u_a(z)
	\end{bmatrix}\in \mathcal{K}_C
	\label{eq34}
\end{equation}
where
\begin{equation}
	\mathcal{K}_C=\{\begin{bmatrix}
		y\\u
	\end{bmatrix}: \begin{bmatrix}
		-\hat{N}_y&\hat{M}_y
	\end{bmatrix}\begin{bmatrix}
		y\\u
	\end{bmatrix}-\hat{M}_yv=0\}
	\label{eq35}\notag
\end{equation} 

Kernel attacks operate within the plant's kernel space, enable them to evada controller-side fault detectors. However, as these attacks are not designed to conform the controller's characteristics, they can be detected by the plant-side detector, underscoring the advantage of plant-side detection for such stealthy attacks.

Furthermore, this stealthy property of kernel attacks serves as an important indicator to distinguish attacks from faults. Kernel attacks only trigger responses in the plant-side detector. In contrast, faults arise from internal malfunctions or sensor anomalies within the plant. Under such conditions, the output under fault, denoted by $y_f$, leads to a control input $u=G_y(y_f)$, which conforms to the controller’s kernel space. As a result, faults are only detected by the controller-side detector. By systematically analyzing the manifestation of abnormal signals across the plant and controller kernel spaces, one can effectively discriminate between faults and stealthy attacks. Based on this key observation, we propose a novel detection framework capable of handling both faults and stealthy attacks.

\begin{figure}[!t]
	\centerline{\includegraphics[width=\columnwidth]{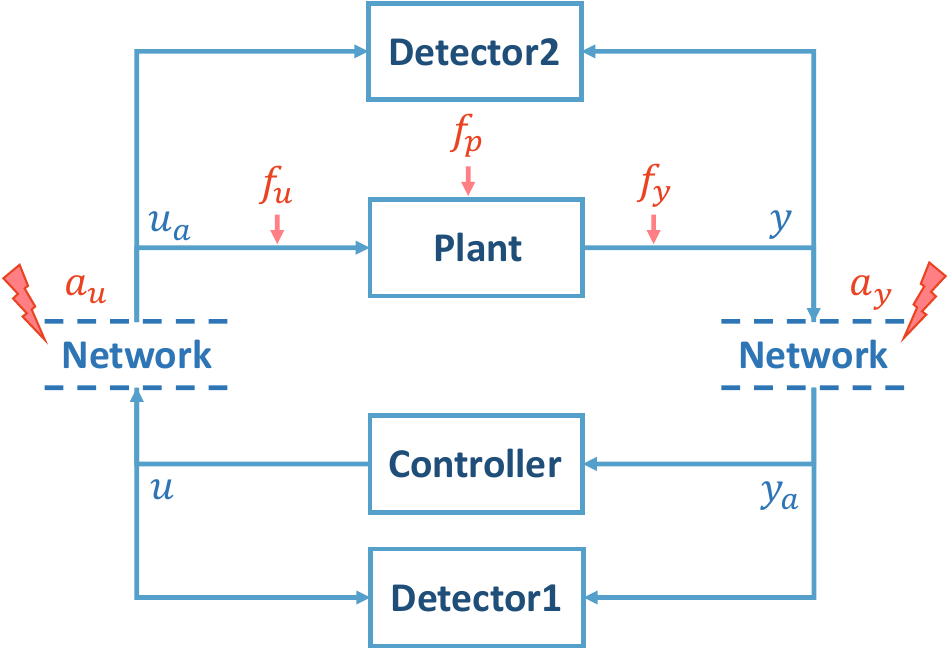}}
	\caption{Dual detection framework.}
	\label{model2}
\end{figure}

As shown in Fig. \ref{model2}, the dual detection framework consists of two anomaly detectors on both sides of the closed-loop dynamic system.  On the controller side, a classic observer-based fault detector (Detector 1) is used to detect system faults. For system \eqref{eq1}, the fault detector is implemented using \eqref{eq2}, \eqref{eq3} and \eqref{eq8}.  On the plant side, a similar input estimator and anomaly detector are deployed. Based on our analysis, this framework enables the detection of both faults and attacks, and facilitates their distinction. The detailed detection strategy is presented in Section \ref{s4}.


$\mathbf{Remark~2.}$ A key aspect of this detection framework is the observability of the control system. Unlike traditional state observers used for control purposes, output reconstructability is sufficient for residual-based detection rather than full state reconstruction\cite{article40}. When the controller system is unobservable, a detectable subsystem can be extracted for detection, or data-driven methods, such as neural networks, can be used to learn data features for controller output reconstruction. This will be addressed in future work.

$\mathbf{Remark~3.}$ 
To exploit the closed-loop characteristics of control systems, some studies have proposed transmitting the control signal $u$ or its encoded form from the plant side to the controller side\cite{article28, article33, article37}. While this approach can improve detection capability, it introduces additional requirements for secure communication channels and potentially increases the information available to attackers. In contrast, our detection framework does not rely on extra channels or encryption transformations, and it can also incorporate such control-signal transmission schemes within its structure.

\subsection{Closed-Loop Stealthiness Condition}\label{s32}
In the dual detection framework, attackers can also design stealthy attacks with system and controller information. Given the plant system \eqref{eq9} and the control system \eqref{eq32} with  detector residuals \eqref{eq19} and \eqref{eq33}. The closed-loop stealthiness condition in the dual detection framework is given by Theorem 1.

$\mathbf{Theorem\ 1}$ Given system model \eqref{eq19}, controller model \eqref{eq33}, and detectors
\eqref{eq15} and \eqref{eq29}, an integrity attack is closed-loop stealthy if and only if conditions \eqref{eq20} and \eqref{eq34} hold.

\textit{proof:} The proof is straightforward and omitted here.

Under the dual detection framework, we define a stealthy attack as closed-loop stealthy attack.

$\mathbf{Definition\ 2}$ An integrity attack is called closed-loop stealthy attack when conditions \eqref{eq20} and \eqref{eq34} hold.

$\mathbf{Remark~4.}$  Compared with the traditional detection scheme, stealthy attacks must satisfy an extra constraint \eqref{eq34}. This constraint reduces the feasible domain of stealthy attack. This is a demanding condition, especially in the feedback control systems.

\section{Detectors Design for closed-loop system}\label{s4}
In this section, we present the implementation of the proposed dual detection framework and provide its optimal design. In addition, we show that this framework is capable of discriminating between faults and attacks by means of the joint decision-making of two detectors.
\subsection{Dual Detector Design}\label{s41}
Considering the system \eqref{eq1} and controller \eqref{eq26}, we design the fault and kernel attack detection scheme under the dual detection framework. For fault detection, detector \eqref{eq3}, \eqref{eq8} is deployed.

The observer-based feedback controller \eqref{eq4} can be rewritten as the following state-space representation
\begin{equation}
	\begin{cases}
		\hat{x}(k+1)=\bar{A}\hat{x}(k)+\bar{B}\bar{y}(k)\\
		u(k)=\bar{F}\hat{x}(k)+\bar{D}\bar{y}(k)+\eta_u(k)
	\end{cases}\label{eq36}
\end{equation}
where $\bar{A}=A+BF-BQC-LC$, $\bar{B}=[L+BQ,B]$, $\bar{y}(k)=[y(k),\bar{v}(k)]^T$, $\bar{F}=F-QC$, $\bar{D}=[Q,I]$, $\eta_u (k)$ is the control signal noise following Gaussian distribution $N(0,\Sigma_{\eta_u })$.

With Kalman filter gain $L_u$, we set an estimator (or a twin) of the controller on the plant side to get the estimate of the control signal $u$.
\begin{equation}
	\begin{cases}
		\hat{x}_u(k+1)=\bar{A}\hat{x}_u(k)+\bar{B}\bar{y}_0(k)+L_u(u(k)-\hat{u}(k))\\
		\hat{u}(k)=\bar{F}\hat{x}_u(k)+\bar{D}\bar{y}_0(k)
	\end{cases}\label{eq37}
\end{equation}
where $\bar{y}_0(k)=[y_0(k),v(k)]=[Cx(k)+Bu(k),v(k)]$. The controller residual can be defined as
\begin{equation}
	r_u(k)=u(k)-\hat{u}(k)\label{eq38}
\end{equation}

Considering the system noise and sensor noise, we analyze the threshold design for attack detector. Let $\epsilon=\hat{x}-\hat{x}_u$. We have
\begin{equation}
	\begin{aligned}
		r(k)&=u(k)-\hat{u}(k)\\
		&=\bar{F}\hat{x}(k)+\bar{D}_1y(k)+\eta_u(k)+\bar{D}_2\bar{v}(k)\\
		&~~-(\bar{F}\hat{x}_u(k)+\bar{D}_1y_0(k)+\bar{D}_2\bar{v}(k))\\
		&=\bar{F}\epsilon(k)+\bar{D}_1\eta(k)+\eta_u(k)
	\end{aligned}\label{eq39}
\end{equation}
From \eqref{eq36}, \eqref{eq37} and \eqref{eq39}, we have
\begin{equation}
	\begin{aligned}
		\epsilon(k)&=\hat{x}(k)-\hat{x}_u(k)\\
		&=\bar{A}\hat{x}(k-1)+\bar{B}_1(y_0(k-1)+\eta(k-1))+\bar{B}_2\bar{v}(k)\\
		&~~-(\bar{A}\hat{x}_u(k-1)+\bar{B}_1y_0(k-1)+\bar{B}_2\bar{v(k)})\\
		&~~-L_u(\bar{F}\eta(k-1)+\bar{D}_1\eta(k-1)+\eta_u(k-1))\\
		&=(\bar{A}-L_u\bar{F})\epsilon(k-1)+\bar{B}_1\eta(k-1)\\
		&~~-L_u(\bar{D}_1\eta(k-1)+\eta_u(k-1))
	\end{aligned}\label{eq40}\notag
\end{equation}
Then, the residual system can be given as
\begin{equation}
	\begin{cases}
		\epsilon(k+1)=(\bar{A}-L_u\bar{F})\epsilon(k)+\bar{\omega}(k)-L_u\bar{\eta}(k)\\
		r_u(k)=\bar{F}\epsilon(k)+\bar{\eta}_u(k)
	\end{cases}\label{eq41}\notag
\end{equation}
where $\bar{\omega}(k)=\bar{B}_1\eta(k)$ and $\bar{\eta}(k)=\bar{D}_1\eta(k)+\eta_u(k)$.

Referring to the property of Kalman filter, we have
\begin{equation}
	r_u(k)\sim N(0,\bar{F}P_u(k)\bar{F}^T+\Sigma_{\bar{\eta}})\notag
\end{equation}
\begin{equation}
	\begin{aligned}
		P_u(k+1)&=(\bar{A}-L_u\bar{F})P_u(k)(\bar{A}-L_u\bar{F})^T+\Sigma_{\bar{\omega}}\\
		&~~-L_u\Sigma_{\bar{\eta}}L_u^T
	\end{aligned}\notag
\end{equation}    

Then, the $\chi^2$ detector with threshold $J_{th,u}$ can be used for kernel attack detection.
\begin{equation}
	J_u(k)=r_u(k)^T\Sigma_{r_u}^{-1}r_u(k)\sim\chi^2(m)\label{eq42}
\end{equation}

The noise covariance matrix and threshold analysis for the controller-side detector are omitted for brevity, since they can be derived in the same manner as for attack detection. The following strategy is employed for detection and discrimination between faults and stealthy  attacks
\begin{equation}
	\begin{cases}
		J(k)>J_{th}, J_u(k)>J_{th,u}&fault~and~kernel~attack\\
		J(k)<J_{th}, J_u(k)>J_{th,u}&kernel~attack\\
		J(k)>J_{th}, J_u(k)<J_{th,u}&fault\\
		J(k)<J_{th}, J_u(k)<J_{th,u}&fault~and~attack~free
	\end{cases}	\label{eq43}\notag
\end{equation}

$\mathbf{Remark~5.}$ When both replay attacks and faults occur simultaneously, the system responds only at the plant-side detector. The replayed data can mask not only the attack behavior but also potential faults due to the open-loop effect, particularly when the system operates stably. To distinguish replay attacks, the defender may combine this framework  with other active detection methods. This issue will be further investigated in future work.

$\mathbf{Remark~6.}$ In practical deployment of the dual detection framework, the plant side generally cannot access the real-time reference input $v$. Therefore, this work focuses on the cases where the reference input satisfies $v(k)=0$ or follows a fixed pattern.

\subsection{Controller Optimization}\label{s42}
If the controller structure is known and fixed, the optimal detection strategy can be implemented following the approach presented in Section \ref{s41}. However, optimal attack detection can be achieved by appropriately selecting the controller parameters.

We first analyze the impact of attacks on the detector.
It is observed that the controller can be expressed as in \eqref{eq17} and \eqref{eq25}. Then, we have
\begin{equation}
	u_0(z)=G_y(z)y(z)=K(z)y(z)\label{eq44}
\end{equation}
From \eqref{eq18}, we set
\begin{equation}
	\begin{cases}
		\hat{M}_y(z)=-X(z)+Q(z)\hat{N}(z)\\
		\hat{N}_y(z)=Y(z)+Q(z)\hat{M}(z)
	\end{cases}\label{eq45}
\end{equation} 
as the coprime factorizations. From the extended Bezout identity
\begin{equation}
	\begin{bmatrix}
		X-Q\hat{N}&Y+Q\hat{M}\\-\hat{N}&\hat{M}
	\end{bmatrix}
	\begin{bmatrix}
		M&-\hat{Y}-MQ\\N&\hat{X}-NQ
	\end{bmatrix}=
	\begin{bmatrix}
		I&O\\O&I
	\end{bmatrix}\label{eq46}\notag
\end{equation}
we can get the Bezout identity for coprime factorization \eqref{eq45}
\begin{equation}
	\begin{bmatrix}
		-\hat{M}&\hat{N}\\-(Y+Q\hat{M})&-X+Q\hat{N}
	\end{bmatrix}
	\begin{bmatrix}
		-\hat{X}+NQ&-N\\ \hat{Y}+MQ&-M
	\end{bmatrix}=
	\begin{bmatrix}
		I&O\\O&I
	\end{bmatrix}\label{eq47}\notag
\end{equation}

Substituting \eqref{eq45} into \eqref{eq33} and in the presence of stealthy attacks, we obtain
\begin{equation}
	\begin{aligned}
		r_{u,a}&=\hat{M}_yu_a-\hat{N}_yy-\hat{M}_yv\\
		&=\hat{N}_ya_y+\hat{M}_ya_u\\
		&=(Y+Q\hat{M})a_y+(-X+Q\hat{N})a_u\\
		&=Ya_y-Xa_u+Q(\hat{M}a_y+\hat{N}a_u)
	\end{aligned}\label{eq48}\notag
\end{equation}
with $R_u(z)=I$. From kernel attack condition \eqref{eq20}, we have 
\begin{equation}
	r_{u,a}=Ya_y-Xa_u\label{eq49}
\end{equation}
From \eqref{eq49}, the effect of a stealthy attack lies within the controller’s kernel space. Moreover, the detection performance is mainly influenced by $X$ and $Y$, which in turn are determined by the controller design variables $F$ and $L$. For fault detection performance, $L$ has been designed with \eqref{eq5}-\eqref{eq7}. Thus, we can design the controller parameter $F$ to improve the attack detection performance. Considering $\mathcal{H}_{-}$ index, the following optimization problem is formulated.
\begin{align}
	\max_{F}~ & \left\| \begin{bmatrix}
		Y&-X
	\end{bmatrix} \right\|_{-}\label{eq50}\\ 
	s.t. ~&\rho(A+BF)<1\notag
\end{align}
where $\|\cdot\|_{-}$ denotes the $\mathcal{H}_{-}$ norm.

Considering the finite-horizon optimization, $X$ and $Y$ can be represented as follows \cite{article29}:
\begin{equation}
	X_{s+n}=\begin{bmatrix}
		-H_{\hat{x},F,L}H_{\hat{x},u,L}&H_{u,F,L}
	\end{bmatrix}\label{eq51}\notag
\end{equation}
\begin{equation}
	Y_{s+n}=\begin{bmatrix}
		-H_{\hat{x},F,L}H_{\hat{x},y,L}&H_{y,F,L}
	\end{bmatrix}\label{eq52}\notag
\end{equation}
\begin{equation}
	H_{\hat{x},L,F}=\begin{bmatrix}
		F^T&(FA_L)^T& \dots &(FA_L^{s-1})^T
	\end{bmatrix}^T\notag
\end{equation}
\begin{equation}
	H_{u,F,L}=\begin{bmatrix}
		I&0&\cdots&0\\-FB_L&I&\ddots&\vdots\\ \vdots&\ddots&\ddots&0\\ -FA_L^{s-2}B_L&\cdots&-FB_L&I
	\end{bmatrix}\notag
\end{equation}
\begin{equation}
	H_{y,F,L}=\begin{bmatrix}
		0&0&\cdots&0\\-FL&0&\ddots&\vdots\\ \vdots&\ddots&\ddots&0\\ -FA_L^{s-2}L&\cdots&-FL&0
	\end{bmatrix}\notag
\end{equation}
\begin{equation}
	H_{\hat{x},u,L}=\begin{bmatrix}
		A_L^{n-1}B_L&\cdots&B_L
	\end{bmatrix}\notag
\end{equation}
\begin{equation}
	H_{\hat{x},y,L}=\begin{bmatrix}
		A_L^{n-1}L&\cdots&L
	\end{bmatrix}\notag
\end{equation}
\begin{equation}
	A_L=A-LC,~B_L=B-LD\notag
\end{equation}
Thus, the optimization problem \eqref{eq50} can be rewritten as a finite-horizon form 
\begin{align}
	\max_{F} ~& \left\| \begin{bmatrix}
		Y_{s+n}&-X_{s+n}
	\end{bmatrix}\right\|_{-}\label{eq53}\\
	s.t. ~&\rho(A+BF)<1\notag
\end{align}
which can be transfer to the optimization problem \eqref{eq54}:
\begin{align}
	\max_{F}&~\gamma\label{eq54}\\
	s.t. &\begin{bmatrix}
		Y_{s+n}&-X_{s+n}
	\end{bmatrix}^T\begin{bmatrix}
		Y_{s+n}&-X_{s+n}
	\end{bmatrix}-\gamma^2 I>0\notag\\
	&~\rho(A+BF)<1\notag
\end{align}

The optimization problem is nonlinear and non-convex. Problem \eqref{eq53} can be solved numerically using nonlinear optimization tools, such as genetic algorithms (GA). Problem \eqref{eq54} can be reformulated as a feasibility problem, and a numerical solution can be obtained using Algorithm \ref{al1}. 
The algorithm starts the search from the largest candidate $\gamma$ to ensure early termination upon finding the maximum feasible value.
\begin{algorithm}[H]
	\caption{Feasibility Search for Optimal Gain $F$}
	\begin{algorithmic}[1]
		\REQUIRE System matrices $A$, $B$; finite-horizon matrices $X_{s+n}$, $Y_{s+n}$;\\
		\hspace{1.9em} search bounds $F_{\min}, F_{\max}$; step size $\Delta F$;\\
		\hspace{1.9em} candidate $\gamma$ range $[\gamma_{\min}, \gamma_{\max}]$ with step size $\Delta \gamma$
		\STATE Initialize $\gamma^* \leftarrow 0$, $F^* \leftarrow$ null
		\FOR{$\gamma = \gamma_{\max}$ to $\gamma_{\min}$ \textbf{step} $-\Delta\gamma$}
		\FOR{each candidate $F$ in $[F_{\min}, F_{\max}]$ with step size $\Delta F$}
		\IF{$\rho(A+BF) < 1$}
		\STATE Compute matrix:
		
		$M = 
		\begin{bmatrix}
			Y_{s+n} & -X_{s+n}
		\end{bmatrix}^\top
		\begin{bmatrix}
			Y_{s+n} & -X_{s+n}
		\end{bmatrix}
		- \gamma^2 I$
		\IF{$M \succ 0$}
		\STATE $\gamma^* \leftarrow \gamma$, $F^* \leftarrow F$
		\STATE \textbf{return} $(F^*, \gamma^*)$
		\ENDIF
		\ENDIF
		\ENDFOR
		\ENDFOR
		\STATE \textbf{return} $(F^*, \gamma^*)$ \quad // If no feasible pair found, $F^* =$ null
	\end{algorithmic}\label{al1}
\end{algorithm}

It is noteworthy that the proposed framework does not compromise fault detection or the implementation of optimal control. When adjusting the matrices $X$ and $Y$, only the controller gain $F$ is modified, while the observer gain $L$ retains its optimal design based on the Kalman filter. This ensures that fault detection performance remains unaffected. Moreover, this adjustment does not degrade control performance. From the perspective of optimal control, once $F$, $L$, and $Q$ have been jointly optimized, any subsequent modification of $F$ can be equivalently compensated by a corresponding adjustment of $Q$, thereby maintaining the same performance as the original joint optimization\cite{article29}. Hence, the proposed design does not impact control performance.

Conventional fault detection approaches do not achieve a true joint design of detection and control, or require compromises in controller performance to enable integration. In contrast, the proposed method leverages the available degrees of freedom in the controller parameters to realize a two-stage joint optimization. The first stage utilizes the controller parameter space to optimize the effect of the attack signal transmission channel, and the second stage applies a unified solution to further improve detection rates. Compared with directly applying the unified solution, this approach achieves a significant performance improvement.

$\mathbf{Remark~7.}$ From residuals \eqref{eq57} and \eqref{eq49}, it can be further shown that the dual detection framework reduces the feasible set of stealthy attacks. For additive attacks, maintaining stealthiness requires satisfying $\hat{M}a_y+\hat{N}a_u=0~\text{and}~Ya_y-Xa_u=0$, which is equivalent to
\begin{equation}
	\begin{bmatrix}
		\hat{N}&\hat{M}\\-X&Y
	\end{bmatrix}\begin{bmatrix}
		a_u\\a_y
	\end{bmatrix}=0.\label{eq59}
\end{equation}
The Bezout identity implies that
\begin{equation}
	rank\left( \begin{bmatrix}X&Y\\ -\hat{N}&\hat{M}\end{bmatrix}\right) =m+p, \notag
\end{equation}
i.e., the matrix is full rank. It then follows that
\begin{equation}
	rank\left( \begin{bmatrix}
		\hat{N}&\hat{M}\\-X&Y
	\end{bmatrix}\right) =m+p, \notag
\end{equation}
which is also full rank, since
\begin{equation}
	\begin{bmatrix}
		\hat{N}&\hat{M}\\-X&Y
	\end{bmatrix}=\begin{bmatrix}
		0&-I\\-I&0
	\end{bmatrix}\begin{bmatrix}
		X&Y\\ -\hat{N}&\hat{M}
	\end{bmatrix}\begin{bmatrix}
		I&0\\0&-I
	\end{bmatrix}. \notag
\end{equation}
Consequently, \eqref{eq59} admits no nonzero solution.  
This indicates that, under these stealthiness conditions, no additive attack can exist.

\begin{figure}[!t]
	\centerline{\includegraphics[width=\columnwidth]{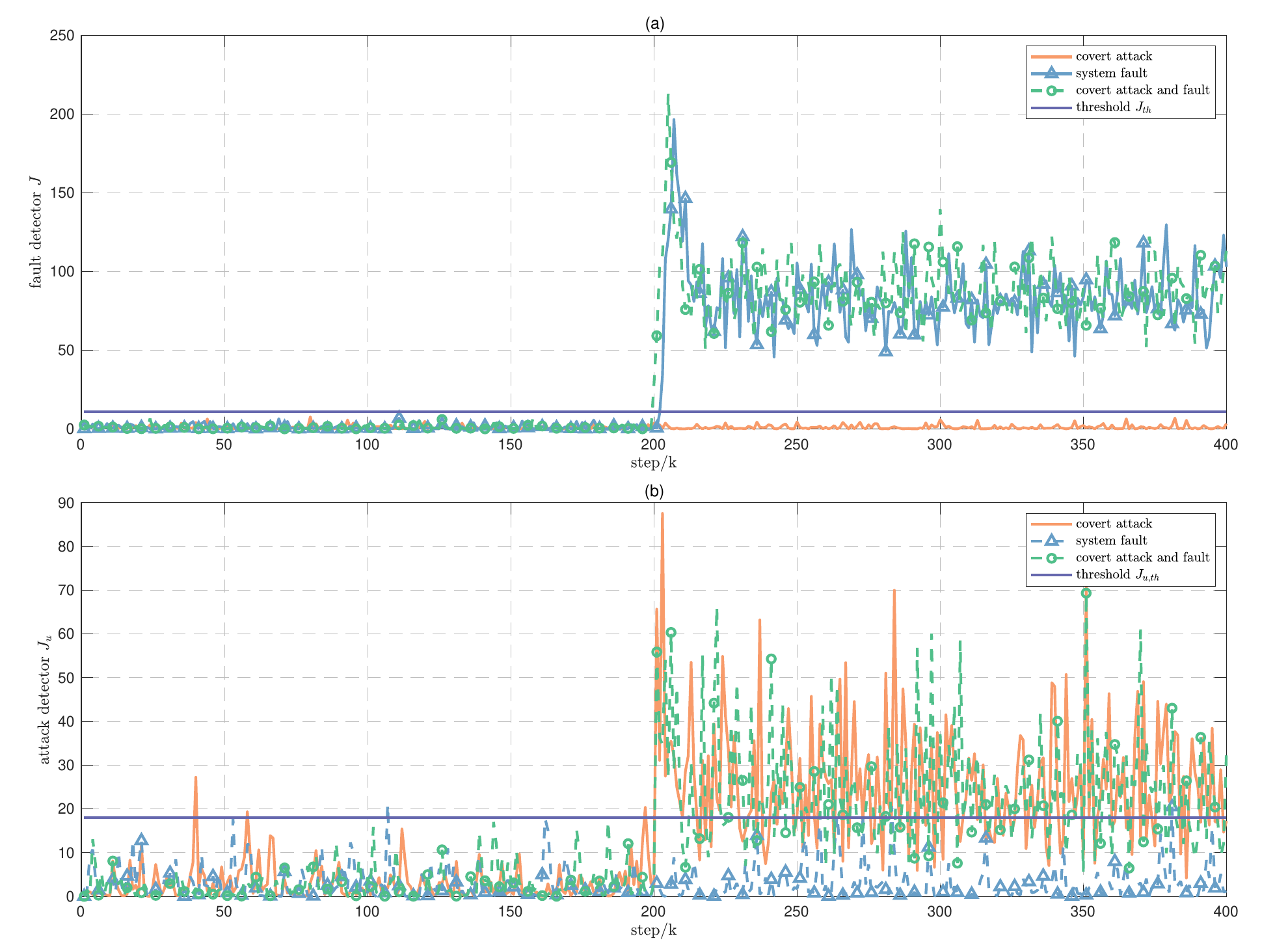}}
	\caption{Detection results to covert attack. (a) fault detector response. (b) attack detector response.}
	\label{covert}
\end{figure}
\begin{figure}[!t]
	\centerline{\includegraphics[width=\columnwidth]{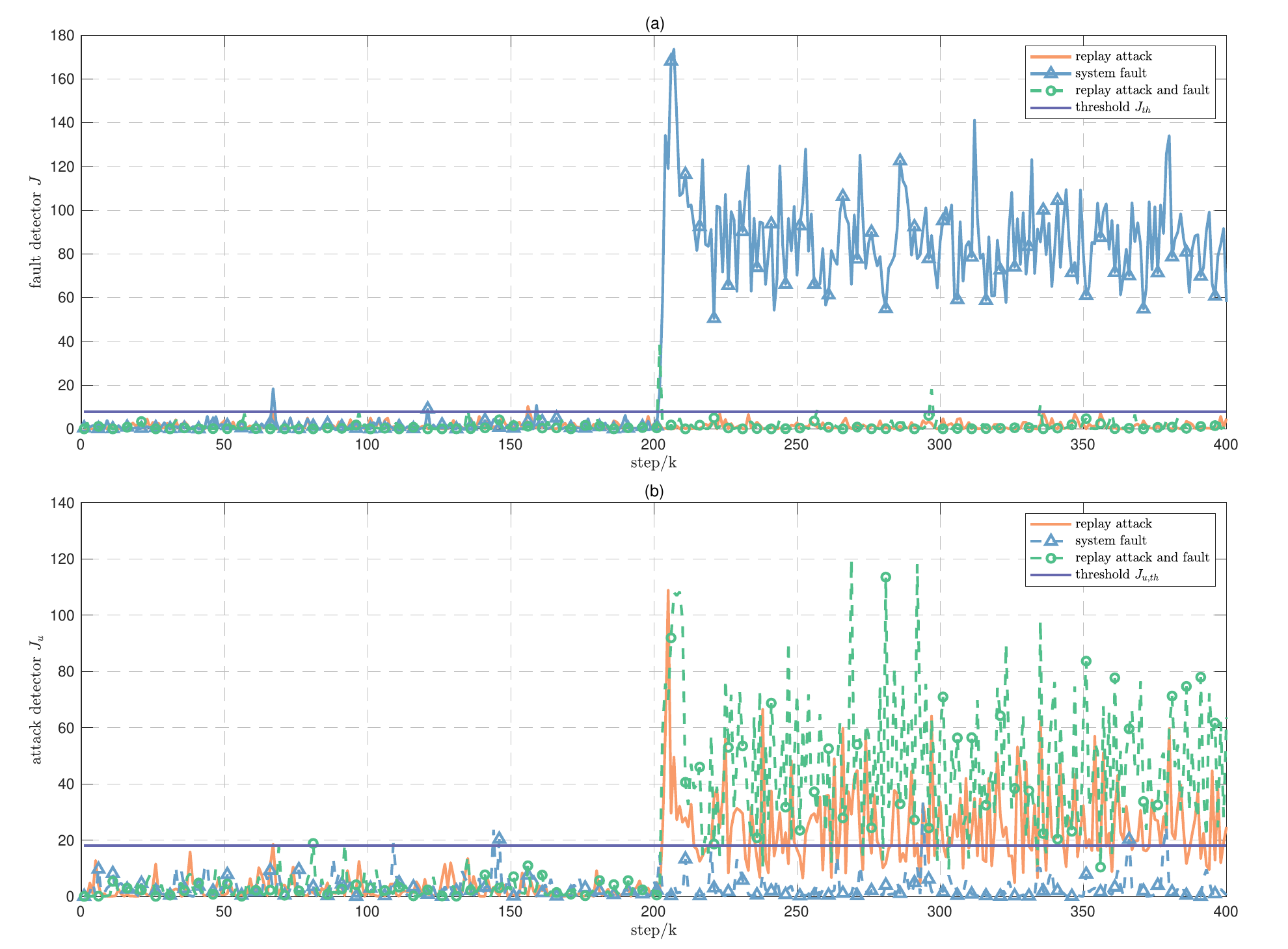}}
	\caption{Detection results to replay attack. (a) fault detector response. (b) attack detector response.}
	\label{replay}
\end{figure}

\section{Simulation study}\label{s5}

In this section, we validate the effectiveness of the proposed dual detection framework through numerical simulations. A linearized longitudinal model of an unmanned aerial vehicle (UAV) is considered\cite{article31}. Under a sampling period of $T_s = 0.1$, the system matrices are given as follows:\\
$A = \begin{bmatrix} 0.8825 & 0.0987 \\ -0.8458 & 0.9122 \end{bmatrix}$, 
$B = \begin{bmatrix} -0.0194 & -0.0036 \\ -1.9290 & -0.3808 \end{bmatrix}$, \\
$C = \begin{bmatrix} 1 & 0 \end{bmatrix}$.

The process noise and sensor noise covariance matrices are set as $\Sigma_w = 0.001I$, $\Sigma_\eta = 0.01I$, and $\Sigma_{\eta_u} = 0.01I$.

A linear quadratic regulator (LQR)  controller is designed by choosing both the state and input weighting matrices as identity matrices. The resulting feedback gain is 
$	F_\text{UAV}= \begin{bmatrix} 0.2550 & -0.3856 ; ~0.0513 & -0.0760 \end{bmatrix}$
\subsection{Detector Response}
This setup is used to demonstrate detection effectiveness under covert attacks and replay attacks. The attack is launched at time step $k_a = 200$, with an attack signal on the control input defined as $a_u = 0.5I$. For the output signal, the covert attack is designed as
$	a_y(k) = \sum_{i=k_a}^{k-1} C A^{k - i-1} B a_u(i)$. 
For replay attacks, the system output is set as $y_a(k) = y(k - k_a)$.
The system fault is modeled as a plant fault with $f_p = 0.5 I$.

The simulation results for the covert attack are shown in Fig.~\ref{covert}. As observed, the covert attack triggers the attack detector while evading the fault detector. In contrast, a system fault only activates the fault detector. When both a fault and a covert attack occur simultaneously, both detectors are triggered.

The simulation results for the replay attack are shown in Fig.~\ref{replay}. Similar to the covert attack, the replay attack can be detected by the attack detector. However, it is worth noting that, since a replay attack opens the control loop, the fault behavior is masked when the fault and the replay attack occur simultaneously. As a result, the fault detector is not triggered in this case.
\begin{figure}[!t]
	\centerline{\includegraphics[width=\columnwidth]{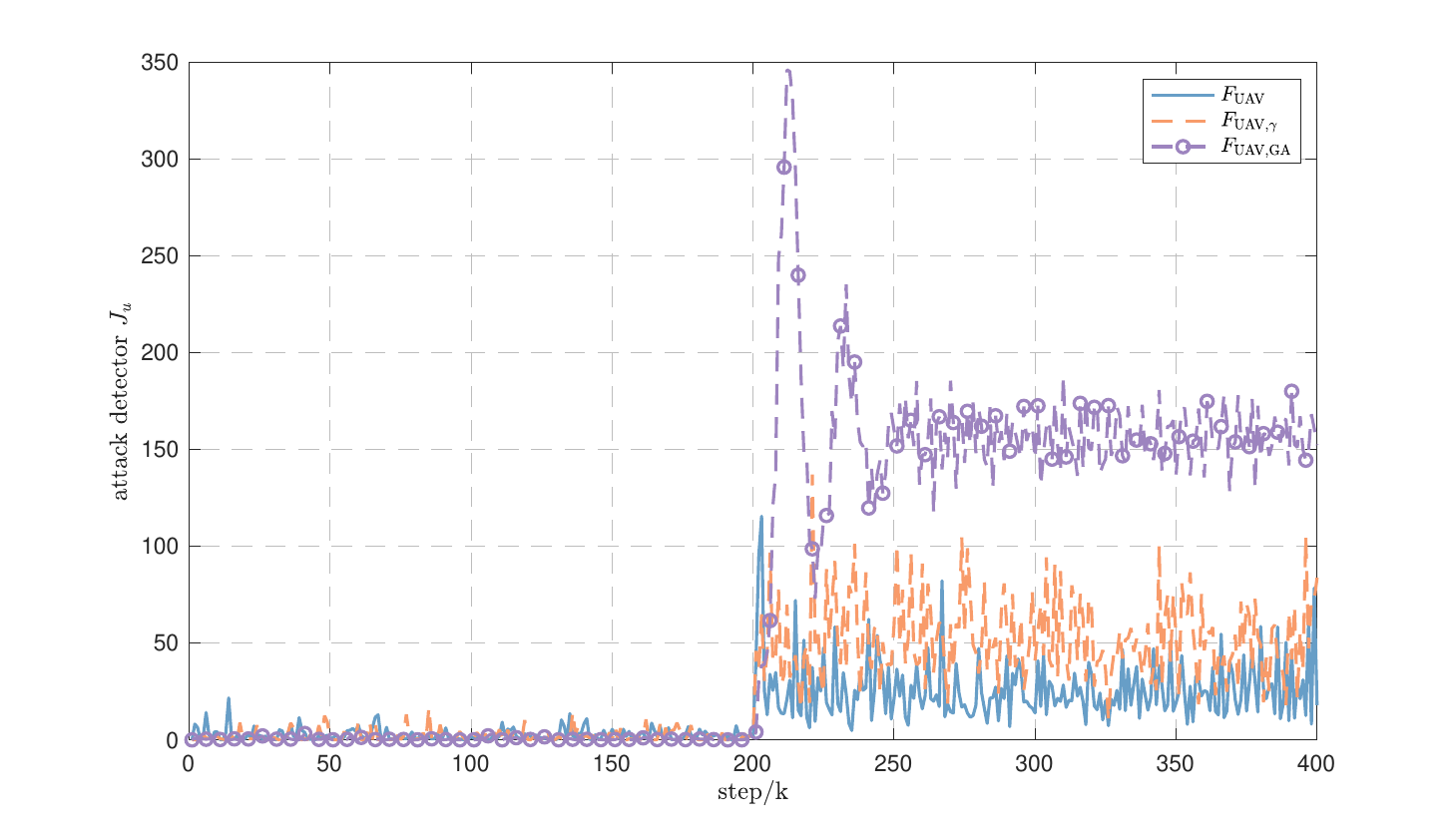}}
	\caption{UAV system detection responses to covert attack.}
	\label{uav}
\end{figure}

\begin{figure}[!t]
	\centerline{\includegraphics[width=\columnwidth]{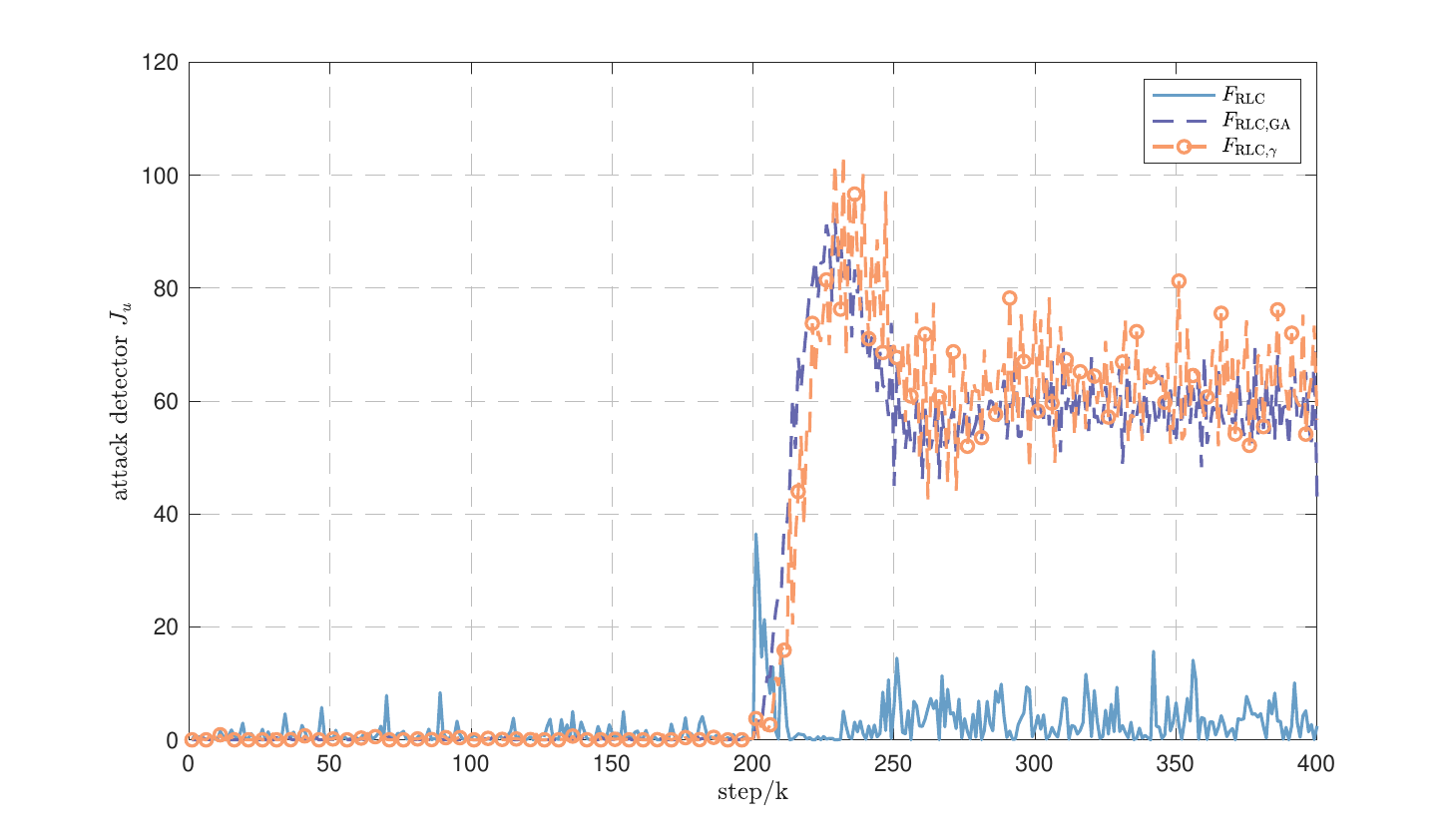}}
	\caption{RLC system detection responses to covert attack.}
	\label{rlc}
\end{figure}
\subsection{Optimization Design}
In this section, we evaluate the effectiveness of optimizing the controller gain matrix  $F$ to improve detection performance. The optimized controller gains $F_{UAV,GA}$ and $F_{UAV,\gamma}$, obtained using GA and Algorithm \ref{al1}, respectively, are presented.

$F_\text{UAV,GA}=[	9.9998,0.4408;9.9996 , 3.7394]$


$F_{\text{UAV},\gamma}=10^{-6} \times[-0.1080 , 0.1050;0.2889, 0.0223]$

The detection results  are shown in Fig.~\ref{uav}. Compared with the LQR controller, both optimized controllers significantly enhance detection performance against kernel attacks, with GA achieving the best performance. This improvement can be attributed to the ability of nonlinear optimization algorithms to explore a wider search space, despite their tendency to fall into a local minima. In practice, different solvers can be employed, and the best-performing controller gain \( F \) can be selected accordingly.

To further verify the generality of the proposed optimization framework, we test it on an RLC system, whose state-space representation is given as follows\cite{article32}:
\begin{equation}
	A = \begin{bmatrix} 1 & -0.1 \\ 0.1 & 0.9 \end{bmatrix}, \quad
	B = \begin{bmatrix} 0.1 \\ 0 \end{bmatrix}, \quad
	C = \begin{bmatrix} 1 & 0 \\ 0 & 1 \end{bmatrix}\notag
\end{equation}

The initial controller is set as $F_{RLC}=[0.8533,-0.1980]$.
All other settings remain the same as in the UAV system. The optimized results are:
\begin{itemize}
	\item GA: $F_{\text{RLC, GA}} = [-10, -10]$
	\item Algorithm 1: 
	$F_{\text{RLC}, \gamma} = [-0.5511, -10]$
\end{itemize}

The corresponding detection results are shown in Fig.~\ref{rlc}. In this case, the feasibility search method outperforms GA. Both optimization methods lead to significant improvements in detection compared with the initial controller gain $F$.

These experiments demonstrate that optimizing the controller gain $F$ can significantly enhance the ability to detect kernel attacks.

\section{Conclusions}\label{s6}
To address the challenges of detecting and distinguishing anomalies in CPSs, this paper proposes a dual detection framework that deploys detectors on both the controller and plant sides. By leveraging information from both ends of the closed-loop system, the proposed framework enhances the detection capability against kernel attacks. 
The proposed framework restricts the feasible set of stealthy attacks to a more stringent condition, defined as the closed-loop stealthiness condition.
Moreover, by combining the outputs of both detectors, the framework enables reliable discrimination between system faults and integrity attacks. 
We further present an implementation and introduce a controller gain optimization strategy to enhance attack detection performance.
The effectiveness of the dual detection mechanism and the benefits of controller optimization are demonstrated through simulation studies.

Future work will extend the proposed framework to nonlinear systems, capturing system dynamics with data-driven methods. Additionally, the joint optimization of observer and controller gains will be explored for detection performance and control objectives.

\bibliographystyle{IEEEtranTIE}
\bibliography{dual_detection}\ 

\end{document}